\documentstyle[12pt]{article}
\newcommand{\fslash}[1]{\mbox{$\!\not\!#1$}}
\newcommand{\Lag}{{\cal L}}

\newcommand{\be}{\begin{equation}}
\newcommand{\ee}{\end{equation}}

\newcommand{\bold}[1]{\mbox{\boldmath ${#1}$}}

\newcommand{\kint}[1]{\int \! \!{d^4 #1 \over {(2\pi)^4}}}

\newcommand{\refb}[1]{(\ref{#1})}

\begin{document}

\baselineskip 4 ex

\title{Quark distributions in the nucleon based on a relativistic
3-body approach to the NJL model}
       
\author{H. Mineo, W. Bentz, K. Yazaki  \\
        Department of Physics, \\
        Faculty of Science \\
        University of Tokyo \\
        Hongo 7-3-1, Bunkyo-ku, Tokyo 113, Japan}

\date{ }
\maketitle
\begin{abstract}
Quark light cone momentum distributions in the nucleon are calculated
in a relativistic 3-body approach to the NJL model by using a simple
'static approximation' for the Faddeev kernel. A method is presented
which automatically satisfies the number and momentum sum rules, even
in the regularized theory. In order to assess the sensitivity to the 
regularization scheme, two schemes which can be formulated in terms of 
light cone variables are discussed. The effects of the
(composite) pion cloud are taken into account in a convolution
approach, and the violation of the Gottfried sum rule is discussed.
After performing the $Q^2$ evolution, the resulting distributions
are compared to the empirical ones.

{\footnotesize PACS numbers: 12.39-x, 12.39.Ki, 14.20.Dh}
\end{abstract}
\footnote{Correspondence to: W. Bentz, E-mail: bentz@phys.s.u-tokyo.
ac.jp}

\newpage

\section{Introduction}
\setcounter{equation}{0}

Deep inelastic lepton-nucleon scattering experiments are providing
detailed information on the quark and gluon distributions in the
nucleon \cite{EXP}. 
By analyzing the distribution functions derived from the 
measured cross sections with the help of the $Q^2$ evolution
based on perturbative QCD, one can extract the fraction of the
nucleon's momentum and spin carried by the quarks at some renormalization
scale $\mu$, 
and obtain valuable information on the spin-flavor structure of the 
nucleon \cite{BOOK}.
For example, of great interest in this connection are the 
flavor dependences of the valence quark distributions  
\cite{FLAV} and  
of the sea quark distributions \cite{STE,KUM}, which reflect
the roles of $qq$ (diquark) and ${\overline q}q$ (mesonic) correlations
in the nucleon. Current experiments at HERA are exploring the
region of small $x$ in order to get more information on large-distance
(non-perturbative) phenomena, and future experiments at RHIC will 
concentrate on the spin distributions.
 
An essential tool to analyze the data on the nucleon structure
functions in the Bjorken limit are the factorization theorems \cite{FAC}, 
which allow us to separate the long-distance parts (parton distributions) 
from the short-distance parts (hard scattering cross sections). 
The latter ones are calculable from perturbative QCD, while the
parton distributions require the knowledge of the nucleon wave
function. The central point of the factorization theorems is that
the infrared-divergent (long-distance) contributions to the perturbative
diagrams can be absorbed into the definition of 'renormalized' parton
distributions \cite{BOOK}. Due to this procedure, the parton distributions
eventually depend on a factorization scale ($\mu$), which is usually
taken to be the same as the renormalization scale. The requirement
that the structure functions should not depend on this scale then
leads to the famous DGLAP equations \cite{DGLAP} for the $\mu$ 
dependence of the
distribution functions, which in turn also determines the $Q^2$
dependence of the structure functions. One can therefore use effective
quark theories to calculate the distribution functions at some
'low energy scale' $\mu=Q_0$, where a description in terms of quark
degrees of freedom alone is expected to be valid, and then use the
DGLAP equations to relate them to the distribution functions extracted
from experimental data at $\mu=Q$.   
  
As an effective quark theory in the low energy region,
the Nambu - Jona-Lasinio (NJL) model \cite{NJL} is a powerful 
tool to investigate
the properties of hadrons \cite{HADR}. It exhibits the spontaneous breaking of
chiral symmetry in a simple and clear way, and allows the solution of 
the relativistic two- and three-body equations in the
ladder approximation due to the simplicity of the interaction 
\cite{FAD1,FAD2}. Concerning the ground state properties of baryons, 
the Faddeev approach to the NJL model has been quite successful \cite{GROUND},
and since in this approach the translational invariance and covariance
are preserved, it is natural to apply it also to the structure functions
of the nucleon.\footnote{There are already several investigations of 
the nucleon
structure functions using the soliton approach to the NJL model \cite{NUCLS}.
For a calculation using the Bethe-Salpeter equation for a quark and a
structureless diquark, see ref. \cite{KUS}.} 
Actually, the NJL model has been
formulated recently on the light cone (LC) and the structure function of the
pion has been calculated \cite{LIGHT}. For the case of the nucleon, 
however, there
is the problem that the relativistic Faddeev equations are usually solved
by performing a Wick rotation using the Euclidean sharp cut-off \cite{FAD1}, 
and this method cannot
be applied directly to the calculation of the LC momentum 
distributions.
A possibility is first to go to the moment space and then re-construct
the distribution functions by the inverse Mellin transformation. 
While such a calculation is now in
progress \cite{NOW}, it is desirable for a first orientation to use 
some simple approximation which
allows a direct calculation of the distributions. The 'static
approximation' to the Faddeev kernel \cite{STAT1,STAT2}, which amounts 
to neglect the
momentum dependence of the propagator of the exchanged quark, allows
an analytic solution of the Faddeev equation, and it has been shown
in ref. \cite{STAT2} that for the nucleon mass this approximation 
is not unreasonable.
However, the overbinding of the nucleon compared to the exact Faddeev
result and the 'point-like' quark-diquark interaction lead to radii which 
are too small \cite{RAD}, and therefore 
we can expect
that the resulting momentum distributions will be too stiff. However, our
main purpose here is to see some general trends, to investigate the 
influence of the regularization scheme and to demonstrate that in
the Faddeev approach the validity of the number and momentum sum rules
is guaranteed from the outset, which is not the case in the soliton or
bag model approaches. The calculation employing the static
approximation to the Faddeev kernel resembles the quark-diquark
calculations performed earlier \cite{DIQ}. We will go beyond these calculations
by including the structure of the diquark (and also of the pion when 
estimating the pionic cloud effects), and investigating the sensitivity
to the regularization scheme while preserving the number and momentum 
sum rules.

The rest of this paper is organized as follows: In sect. 2 we explain
the model for the nucleon wave function, in sect. 3 we explain our
method to calculate the quark distribution functions, and in sect. 4
we discuss the numerical results. A summary is presented in sect. 5.

\section{NJL model for the nucleon wave function}
\setcounter{equation}{0}

The NJL model is characterized by a chirally symmetric four-Fermi
interaction lagrangian $\Lag_I$. By means of Fierz transformations, one
can rewrite any $\Lag_I$ into a form where the interaction strength
in a particular ${\overline q}q$ or $qq$ channel can be read off directly
\cite{FAD1}.
That part which generates the constituent quark mass $M_Q$ and the pion
as a collective ${\overline q}q$ bound state is given by
\be
{\cal L}_{I,\pi} = \frac{1}{2}g_{\pi}
\biggl( \Bigl(\overline{\psi}\psi \Bigr)^2
	- \Bigl(\overline{\psi}(\gamma_5\bold{\tau})\psi\Bigr)^2
\biggr), \label{lagpi}
\ee
and that which describes the $qq$ interaction in the scalar diquark
($J^{\pi}=0^+, T=0$) channel is
\be
{\cal L}_{I,s} = g_s \left(\overline{\psi}\left(\gamma_5 C\right)\tau_2
\beta^A \overline{\psi}^T\right) \left(\psi^T\left(C^{-1}\gamma_5\right)
\tau_2 \beta^A \psi\right),
\label{lags}
\ee
where $\beta^A=\sqrt{3/2}\,\, \lambda^A \,\,(A=2,5,7)$ are the color 
${\overline 3}$ matrices, and $C=i\gamma_2 \gamma_0$. The coupling
constants $g_{\pi}$ and $g_s$ are related to the ones appearing in
the original $\Lag_I$ by numerical factors due to the Fierz transformation, 
but instead of choosing
a particular form of $\Lag_I$ we will treat $g_{\pi}$ and the ratio
$r_s=g_s/g_{\pi}$ as
free parameters, where the latter reflects different possible forms 
of $\Lag_I$ \cite{FAD1}.

The reduced t-matrices in the pionic and scalar diquark channels are
obtained from the respective Bethe-Salpeter (BS) equations as \cite{FAD1}
\be
\tau_{\pi}(k)={-2ig_{\pi} \over 1 + 2 g_{\pi} \Pi_{\pi}(k)}, \,\,\,\,\,\,\,
\,\,\,\,\,\,\,\, 
\tau_{s}(k)={4ig_s \over 1 + 2 g_s \Pi_s(k)}
\label{tau} 
\ee
with the bubble graph
\be
\Pi_{\pi}(k)=\Pi_s(k)=6 i \kint{q} tr_D \biggl[
	\gamma_5 S(q) \gamma_5  S(q-k)\biggr],
\label{bubb}
\ee
where ${\displaystyle S(q) = {1 \over \fslash{q} - M_Q +i\epsilon}}$ is 
the Feynman propagator and $M_Q$ the constituent quark mass.

If the interacting two-body channels are restricted to the scalar
diquark one, the relativistic Faddeev equation \cite{FAD1} can be reduced to
an effective BS equation for a composite scalar diquark and a quark,
interacting
via quark exchange \cite{CAH}. As we explained in the previous section, in this
paper we will restrict ourselves to the static approximation 
\cite{STAT1,STAT2}, where 
the Feynman propagator in the quark exchange kernel
is simply replaced by $-1/M_Q$. Then the effective BS
equation reduces to a geometric series of quark-diquark bubble graphs
($\Pi_N(p)$), and the solution for the t-matrix in the color singlet
channel is
\be
T(p)=\frac{3}{M_Q} \frac{1}{1+\frac{3}{M_Q} \Pi_N(p)}
\ee
with
\be
\Pi_N(p)=-\kint{k} S(k) \tau_s(p-k).
\ee
The quark-diquark vertex function $\Gamma_N(p)$ in the covariant 
normalization is then obtained
from the pole behaviour $T\rightarrow \sum_s \Gamma_N(ps){\overline
\Gamma}_N(ps)/(p^2-M_N^2+i\epsilon)$, where $M_N$ is the nucleon mass, as
\footnote{Our conventions for LC
variables are $a^{\pm}=\frac{1}{\sqrt{2}}\left(a^0\pm a^3\right),\,\,
a_{\pm}=\frac{1}{\sqrt{2}}\left(a_0\pm a_3\right),$ and $a_{\perp}^i=
-a_{\perp i}\,\,(i=1,2)$. The Lorentz scalar product is
$a\cdot b=a_+ b^+ + a_- b^- - {\bold a}_{\perp}\cdot {\bold b}_{\perp}$.
We will frequently call $p_- (p_+)$ the 'LC minus (plus) component' of the
four vector $p$.}
\begin{eqnarray}
\Gamma_N(p)&=& Z_N u_N(p) \label{vert}
\\
Z_N &=& \left(			
	\frac{1}{\left. {\partial \Pi_N(p)/ \partial \fslash{p} }\,\,
	\right|_{\fslash{p}=M_N }
	}\right)^{\frac{1}{2}}
	= \left(
	\frac{p_-/M_N }
	{ {\bar u}_N(p) \frac{\partial \Pi_N(p)}{\partial p_+ } u_N(p) }
\right)
           ^{\frac{1}{2}},
\label{zn}
\end{eqnarray}
where $u_N(p)$ is a free Dirac spinor with mass $M_N$ normalized by
${\overline u}_N(p)\,u_N(p)=2M_N$.
In this normalization, the vertex function satisfies the relation
\be
\frac{1}{2 p_-}\,\, {\overline \Gamma}_N(p) \frac{\partial \Pi_N(p)}
{\partial p_+}\,\, \Gamma_N(p) = 1,
\label{curr}
\ee
which leads to charge and baryon number conservation in any treatment
which preserves the Ward identity
\be
\Lambda_{q/P}^+(p,p)=\frac{\partial \Pi_N(p)}{\partial p_+}\,\, N_{q/P}
\label{ward1}
\ee
for the vertex of the quark number current of the proton at $q=0$ 
($N_{u/P}=2,\,\,N_{d/P}=1$).
\footnote{This vertex is defined by $\langle p|{\overline \psi}(0) \gamma^+
\frac{1\pm \tau_z}{2}
\psi(0)|p\rangle = {\overline \Gamma}_N(p) \Lambda_{q/P}^+(p,p) \Gamma_N(p)$
for $q=u(d)$.}

 

\section{Quark distribution functions}
\setcounter{equation}{0}

The twist-2 quark LC momentum distribution in the proton
(momentum $p$) is defined as \cite{JAFFE,JAFFE1}
\be
\tilde{f}_{q/P}(x) = \frac12 \int\frac{dz^- }{2\pi}e^{ip_{-}xz^- } \langle 
p\vert T\left({\overline \psi_q}(0)\gamma^+ \psi_q(z^- )\right)\vert p
\rangle, 
\label{dis}
\ee
where $q$ denotes the quark flavor, $|p\rangle$ is the proton state, 
and $x$ is the Bjorken variable
which corresponds to the fraction of the proton's LC momentum component
$p_-$ carried by the quark $q$. 
As has been discussed in detail in ref. 
\cite{JAFFE,JAFFE1}, for connected LC correlation functions the T-product
is identical to the usual product, from which it follows that
the distribution \refb{dis} is non-zero in the interval $-1<x<1$.
The physical quark and antiquark distributions which determine
the structure functions $F_1$ and $F_2$ are obtained for $0<x<1$
as $f_{q/P}(x)=\tilde{f}_{q/P}(x)$ and $f_{{\overline q}/P}(x)
=-\tilde{f}_{q/P}(-x)$. The valence
($v$) and sea ($s$) quark distributions are then given by 
$f_{q_v/P}(x)=f_{q/P}(x)-f_{{\overline q}/P}(x)$, 
$f_{q_s/P}(x)=f_{{\overline q}_s/P}(x)=f_{{\overline q}/P}(x)$.

The evaluation of the distribution \refb{dis} can be reduced to a
straight forward Feynman diagram calculation by noting that it can
be expressed as \cite{JAFFE,JAFFE1,KPW}
\footnote{The original Lorentz invariant expression is recovered by
$k_-/p_- \rightarrow k\cdot q /p\cdot q$ and $\gamma^+/p_- \rightarrow
\fslash{q}/p\cdot q$. The expression \refb{opdef} corresponds to the
Bjorken limit in the frame where ${\bold q}_{\perp}=0,\,\,q_z <0$,
i.e; $q_+\rightarrow \infty,\,\,q_-\rightarrow -p_- x$.} 
\be
\tilde{f}_{q/P}(x)=-\frac{i}{2p_-} \kint{k} \,\,\delta(x-\frac{k_-}{p_-})
\,\, Tr_D \,\,\gamma^+ M_q(p,k),
\label{opdef}
\ee
with the quark 2-point function in the proton given by
\be
M_{q,\beta \alpha}(p,k) = i \int d^4 z\,\,e^{ikz}\,\, 
<p|T\left({\overline \psi}_{q,\alpha}
(0) \psi_{q,\beta}(z)\right)|p>.
\ee
We therefore have to evaluate the Feynman diagrams for the quark propagator 
in the nucleon, trace it with $\gamma^+$,
fix the LC minus component of the quark momentum as 
$k_-=p_- x$ and integrate
over the remaining components $k_+$ and ${\bold k}_{\perp}$. 
Since in our model for the nucleon discussed in the previous section
the quark can either appear as a spectator or as a constituent of the
scalar diquark, the Feynman diagrams to be evaluated are shown in
fig.1. (In the full Faddeev approach, there is also a diagram where the
external operator acts on the exchanged quark, but in the present static
approximation this diagram does not contribute.) 

To present the formulae for the diagrams fig.1, we note that the
second diagram (the diquark contribution) can be expressed
conveniently as a convolution integral if we insert the identity
\be
1=\int dy \int dz \int dq_0^2 \,\,\delta(y-\frac{q_-}{p_-})\,\, 
\delta(z-\frac{k_-}{q_-})\,\, \delta(q^2-q_0^2),
\ee 
i.e; $y$ is the fraction of the nucleon's momentum component $p_-$ 
carried by the diquark, $z$ is the fraction of the diquark's momentum
component $q_-$ carried by the quark inside the diquark ($x=yz$), and
$q_0^2$ is the virtuality of the diquark. Using also the identity 
\be
S(k) \gamma^+ S(k) = -\frac{\partial S(k)}{\partial k_+},
\label{ward}
\ee
and performing partial integrations in the plus components of the loop
momenta, which is permissible since these integrations are convergent and 
not restricted by the regularization schemes to be discussed later, 
we obtain the following expression\footnote{To distinguish the
case without the pion cloud (valence quark picture) from that including
the pion cloud, we replace $q\rightarrow Q$ in the formulae corresponding
to the diagrams of fig.1.}:
\be
f_{Q/P}(x)=\delta_{Q,U}\,\,F_{Q/P}(x) + \frac{1}{2} F_{Q(D)/P}(x).
\label{gen}
\ee
Here the first term corresponds to the first diagram in fig.1 and is
expressed as
\be
F_{Q/P}(x) = \frac{1}{2p_- } {\bar \Gamma_N }\left( \frac{\partial }
{\partial p_+}
	\Pi_N (x,p) \right) \Gamma_N,  
\label{fq}
\ee
where $\Pi_N(x,p)$ is the quark-diquark bubble graph for fixed minus
component of the quark momentum:
\be
\Pi_N (x,p) = -\int \frac{d^4 k}{(2\pi)^4 }\delta ( x-\frac{k_- }{p_- }
	)S(k) \tau_s (p-k). 
\label{piq}
\ee
The second term in \refb{gen} corresponding to the second diagram
in fig.1 is purely isoscalar and given by the convolution integral 
\be
F_{Q(D)/P}(x)=\int_0^1 dy \int_0^1 dz \delta (x-yz) 
\int_{-\infty}^{\infty}
	d q_0^2 \,\,F_{Q/D} (z,q_0^2)\,\,F_{D/P}(y,q_0^2),
\label{conv}
\ee
where the distributions $F_{Q/D}(z,q_0^2)$ and $F_{D/P} (y,q_0^2)$ 
for fixed virtuality of the diquark (D) 
are expressed
as
\begin{eqnarray}
F_{Q/D} (z,q_0^2 ) &=& -2g^2 (q_0^2 )\,\, \frac{\partial \Pi_s (z,q_0^2 )}{
	\partial q_0^2 }
\label{fqd} \\
F_{D/P} (y,q_0^2 ) &=& {\bar \Gamma_N } \left( 
	\frac{1}{2p_- }\frac{\partial }{\partial p_+ }
		+y\frac{\partial }{\partial q_0^2 }
	\right)\Pi_N (y,q_0^2 ;p)\,\,\Gamma_N.
\label{fd}
\end{eqnarray}
Here $\Pi_s(z,q_0^2)$ is the quark-quark bubble graph for fixed minus
momentum component of the quark, 
${\displaystyle g^2(q_0^2)=-1/\left(\frac{\partial \Pi_s(q_0^2)}
{\partial q_0^2}\right)}$
is the quark-diquark coupling constant, and $\Pi_N(y,q_0^2;p)$ is the
quark-diquark bubble graph for fixed virtuality and minus momentum
component of the diquark:
\begin{eqnarray} 
\Pi_s(z,q_0^2) &=& \left[6 i \kint{k} \delta(z-\frac{k_-}{q_-}) tr_D \left(
	\gamma_5 S(k) \gamma_5  S(k-q)\right)\right]_{q^2=q_0^2}
\label{pisq} \\
\Pi_N (y,q_0^2 ; p) &=& -\int \frac{d^4 q}{(2\pi)^4 }\delta (y-
	\frac{q_- }{p_- } ) \delta(q^2 -q_0^2 )S(p-q)\tau_s (q).
\label{pid}
\end{eqnarray}
If we use the dispersion 
representation
for the diquark t-matrix $\tau_s$, we can perform the $k_+$ and $q_+$
integrals analytically and verify that the distribution \refb{gen}
is non-zero only in the interval $0<x<1$. This fact was anticipated
already
in our notation ($\tilde{f}_{Q/P}(x)=f_{Q/P}(x)$) and in the integration
limits in eq. \refb{conv}, and corresponds to
a valence quark model ($f_{{\overline Q}/P}(x)=0$).  

Using the above expressions and the normalization \refb{curr}, it is a
simple matter to confirm the validity of the number and momentum sum rules
\footnote{To verify these relations, it is only necessary to note 
that the relations
${\displaystyle \int_0^1 dz F_{Q/D}(z,q_0^2)=2}$ and
${\displaystyle \int_0^1 dz\cdot z F_{Q/D}(z,q_0^2)=1}$ hold for
any $q_0^2$ (here $F_{Q/D}(z,q_0^2)$ is symmetric around $z=1/2$), and 
therefore the second term in eq. \refb{fd} gives a vanishing surface term
when integrated over $q_0^2$. The sum rules are then obvious since
the integral of $\Pi_N(y,q_0^2,p)$ over $q_0^2$
reduces to $\Pi_N(1-y,p)$, and that of $\Pi_N(x,p)$ over $x$ to $\Pi_N(p)$.}
\begin{eqnarray} 
\int_0^1\,\,dx\,\, f_{Q/P}(x) &=& N_{Q/P} \label{nr1} \\
\int_0^1\,\,dx\cdot x \left(f_{U/P}(x)+f_{D/P}(x)\right) &=& 1.
\label{mom1} 
\end{eqnarray}
In more general terms, what we have really
confirmed here is the validity of the Ward identities for quark number
and momentum conservation: Concerning the number conservation,  
if we integrate the distribution function of fig.1 over $x$,
the restriction $k_-=p_- x$ is lifted, and the diagrams correspond 
to $\Lambda^+_{Q/P}(p,p)/2p_-$. The validity of the Ward identity \refb{ward1} 
follows then
from \refb{ward} and partial integrations in the plus
components of the loop momenta. A similar argument holds for the Ward
identity expressing momentum conservation. 
Therefore, the Ward identities and the sum rules \refb{nr1}, \refb{mom1}
hold in any regularization scheme which does not restrict the LC 
plus components of the loop momenta. The regularization schemes 
to be discussed at the end of this section satisfy this requirement. 

As we have noted above, the model described so far gives essentially
only valence-like distributions at the low energy scale. Although sea 
quark distributions will be generated in the process of the $Q^2$
evolution, those will be flavor independent ($f_{{\overline u}/P}=
f_{{\overline d}/P}$),
which contradicts the experimentally measured violation of the
Gottfried sum rule \cite{EXP}. Also, it is a general trend of valence 
quark models
that the resulting valence quark distributions are too stiff (too strongly
pronounced peak and too small
support at low values of $x$). We therefore consider here the effects
of the pion dressing of the constituent quarks, as has been done also
in previous works \cite{PION}. (For a recent investigation using the
LC quantization, see ref. \cite{FZ}.) In order to take into account 
the pion cloud, 
in principle we should solve the Schwinger-Dyson equation for the 
quark Feynman propagator
${\displaystyle S(q) = {1 \over \fslash{q} - M_Q - \Sigma_Q(p)}}$,
where
\be
\Sigma_Q(p)=-3 \kint{k}\left(\gamma_5 S(k) \gamma_5 \right)  
\tilde \tau_{\pi}(p-k) \equiv -\Pi_Q(p)
\label{self}
\ee
is the quark self energy due to the pion cloud, and the reduced pion 
t-matrix $\tilde \tau_{\pi} \equiv \tau_{\pi}+2ig_{\pi}$
depends also on $S(p)$. This propagator should then be used to
calculate the diquark t-matrix $\tau_s$ and the nucleon vertex function
$\Gamma_N$. Using
these modified propagators and vertex functions, one should, in addition
to the diagrams of fig.1, also evaluate the diagrams of fig.2, where
the operator insertion is made on the quark while the pion is 
'in flight', or on the quark and the antiquark in the pion. 

Clearly, such a calculation is very complicated, and the usually employed
convolution formalism \cite{PION} involves the following two major 
approximations:
First, pionic effects can be renormalized into a re-definition of
the constituent quark mass and the four fermi coupling constants if
one approximates the quark propagator by its pole part
\footnote{Green functions and vertex functions
which differ from
those without pionic effects by the replacements $M_Q\rightarrow \hat{M}_Q$,
$G_{\alpha}\rightarrow \hat{G}_{\alpha}\equiv G_{\alpha} Z_Q^2
\,\,(\alpha=\pi,s)$ will be denoted by a hat. Since due to the discussion
following eq.\refb{zq} these renormalized quantities are numerically
equivalent to the ones used previously, this distinction will eventually
be dropped.}
\be
S(p)=\frac{Z_Q}{\fslash{p}-\hat{M}_Q+i \epsilon} \equiv 
Z_Q \hat{S}(p)
\ee
with 
\be
Z_Q=\left(1+\frac{\partial \Pi_Q}{\partial \fslash{k}}|
_{\fslash{k}=\hat{M}_Q}\right)^{-1}.
\label{zq}
\ee
If we define 'renormalized' coupling constants $\hat{G}_{\alpha}$
by $G_{\alpha}=\hat{G}_{\alpha}/Z_Q^2 \,\,(\alpha=\pi,s)$, it is
easy to see that the Green functions and vertex functions are
renormalized according to $\tau_{\alpha}=\hat{\tau}_{\alpha}/Z_Q^2,
\,\,T=Z_Q\,\,\hat{T}, \Gamma_N=\sqrt{Z_Q}\hat{\Gamma}_N$. If we
then impose the same conditions on the parameters as commonly used in the case
without pion cloud effects (that is, $f_{\pi}=93 MeV$, $m_{\pi}=140 MeV$,
$\hat{M}_Q=300 - 500 MeV$, and $M_N=940 MeV$), the cut-off and the coupling
constants $\hat G_{\alpha}$ take the same values
as in the case without pionic cloud effects. (In terms of the lagrangian,
such a renormalization procedure corresponds to
writing $G\left({\overline \psi}\psi\right)^2=\hat{G}
\left( \overline{\hat{\psi}}\hat{\psi}\right)^2$ with 
$\psi=\sqrt{Z_Q}\,\,\hat{\psi}$ and $G=\hat{G}/Z_Q^2$.) It is
also easy to check that all diagrams in fig.1 and fig.2 get a factor
$Z_Q$. 

If one then calculates the diagrams of fig.2 in terms of these
renormalized quantities and writes the results in terms of a
convolution integral, one finds that due to the Dirac structure
of the insertions on the 'parent' quark line there appear three
convolutions terms \cite{KUL}. Only one of them involves the generalization
of the 'bare' quark distribution in the nucleon (eq.\refb{gen}) to the
off shell case ($f_{Q/P}(x)\rightarrow f_{Q/P}(x,k_0^2)$, where 
$k_0^2$ is the virtuality
of the parent quark), convoluted with the quark distribution within the 
parent quark. Each of the other two terms involve one additional
factor of $(k_0^2-M_Q^2)$ in the integrand compared to the first term, which
has a sharp peak at $k_0^2=M_Q^2$. The second approximation commonly used is
therefore to neglect these two terms, and to assume that due to the sharp
peak of $f_{Q/P}(x,k_0^2)$ the quark distribution within the parent
quark can be evaluated at $k_0^2=M_Q^2$ and taken outside of the $k_0^2$
integral \cite{JAFFE}. In this way one arrives at the familiar convolution 
form
\be
f_{q/P}(x) = \sum_{Q=U,D} \int_0^1 dy \int_0^1 dz \delta(x-yz) 
f_{q/Q}(z) f_{Q/P}(y)
\label{convo}
\ee
and a similar expression with $q\rightarrow {\overline q}$,
where the parent quark distribution in the proton $f_{Q/P}$ is given
by eq.\refb{gen}, and $f_{q/Q}$ ($f_{{\overline q}/Q}$) is the quark 
(antiquark) distribution within an
on-shell parent quark which is obtained by evaluating the Feynman diagrams
shown in fig.3. 

The quark and antiquark distributions in the parent quark obtained from 
the diagrams of fig.3 can be expressed 
as
\begin{eqnarray}
f_{u/U}(x)&=&Z_Q\delta(x-1) + \frac{1}{3} F_{q/Q}(x) + \frac{5}{6}
F_{q(\pi)/Q}(x) 
\label{fup} \\
f_{d/U}(x)&=& \frac{2}{3} F_{q/Q}(x) + \frac{1}{6}
F_{q(\pi)/Q}(x) 
\label{fdp} \\
f_{{\overline u}/U}(x)&=& \frac{1}{6} F_{q(\pi)/Q}(x) 
\label{fau} \\
f_{{\overline d}/U}(x)&=& \frac{5}{6} F_{q(\pi)/Q}(x).
\end{eqnarray}
The distributions for $Q=D$ are also determined from these expressions
due to isospin symmetry. The detailed formulae for the distributions 
$F_{q/Q}(x)$ and 
\be
F_{q(\pi)/Q}(x)=\int_0^1 dy \int_0^1 dz \delta (x-yz) \int_{-\infty}
^{\infty} d q_0^2 \,\,F_{q/\pi} (z,q_0^2)\,\,F_{\pi/Q}(y,q_0^2),
\label{fpi}
\ee
corresponding to the second and third diagrams of fig.3, respectively, 
can be obtained
from the previous expressions \refb{fq}-\refb{pid} 
as follows: $F_{q/Q}(x), F_{q/\pi}$ and
$F_{\pi/Q}$ are given by expressions similar to eqs.\refb{fq}, \refb{fqd} and
\refb{fd}, respectively, but with the
following replacements: (i) The nucleon spinor $\Gamma_N$ 
is replaced by the quark spinor defined via the residue of the propagator
$S(p)$: $\Gamma_Q(p)=Z_Q\,\,u_Q(p)$, 
where $u_Q(p)$ is a free Dirac spinor with mass $M_Q$ normalized by
${\overline u}_Q(p)\,u_Q(p)=2M_Q$. From eq.\refb{zq} it follows that
this spinor satisfies the relation
\be
\frac{1}{2 p_-}\,\, {\overline \Gamma}_Q(p) \frac{\partial \Pi_Q(p)}
{\partial p_+}\,\, \Gamma_Q(p) = 1-Z_Q.
\label{curr1}
\ee
(ii) the polarizations $\Pi_N(x,p)$ and
$\Pi_N(y,q_0^2;p)$ are replaced by $\Pi_Q(x,p)$ and
$\Pi_Q(y,q_0^2;p)$. These are defined analogously to eqs. \refb{piq} and
\refb{pid} by introducing
the $\delta$ function insertions to fix the minus momentum components
of the quark or the pion and the virtuality $q_0^2$ of the pion 
into $\Pi_Q$ defined by eq. \refb{self} instead
of $\Pi_N$, and (iii) $F_{q/\pi}$ is given by the r.h.s. of 
eq. \refb{fqd}, \
but without the overall factor 2. We note that $f_{q/Q}$, and therefore also
the distribution $f_{q/P}$ of eq. \refb{convo}, involves an overall factor
$Z_Q$, in accordance with our discussion following eq. \refb{zq}.

>From these expressions and the normalization \refb{curr1}, the validity 
of the number and momentum sum rules
\begin{eqnarray}
\lefteqn{\int_0^1 dx \left(f_{q/Q}(x)-f_{{\overline q}/Q}(x)\right)=
\delta_{Q,q}}
\label{nr2} \\
& &\int_0^1 dx\cdot x \left(f_{u/Q}(x)+f_{{\overline u}/Q}(x)
+f_{d/Q}(x)+f_{{\overline d}/Q}(x) \right)= 1
\label{mom2}
\end{eqnarray}
can be easily checked in the same way as outlined above for the parent quark
distributions in the proton. (The number sum rule is a consequence of the 
Ward identity for the quark number current of the parent quark
${\displaystyle \Lambda_{q/Q}^+(p,p)=\frac{\partial \Pi_Q(p)}{\partial p_+} 
\delta_{q,Q}}$.) The validity of the number and momentum sum rules
\begin{eqnarray} 
\int_0^1 dx\,\,f_{q_v/P}(x) \equiv \int_0^1
 dx \,\,\left(f_{q/P}(x)
-f_{{\overline q}/P}(x)\right) &=& N_{q/P}
\label{nr3} \\
\int_0^1 dx \cdot x \,\,\left(f_{u/P}(x)+f_{{\overline u}/P}(x)
+f_{d/P}(x)+f_{{\overline d}/P}(x) \right) &=& 1
\label{mom3}
\end{eqnarray}
is then a consequence of eqs. \refb{nr1}, \refb{mom1}, \refb{nr2}, 
refb{mom2} and \refb{convo}.

Of particular interest is also the Gottfried sum
\begin{eqnarray}
S_G &=& \frac{1}{3} \int_0^1 dx\,\,\left(f_{u/P}(x)+f_{{\overline u}/P}(x)-
f_{d/P}(x)-f_{{\overline d}/P}(x)\right) \nonumber \\
&=& \frac{1}{3} - \frac{4}{9}
\int_0^1 dx\,\,F_{q/Q}(x) = \frac{1}{3} - \frac{4}{9} \left(1-Z_Q\right),
\label{gs}
\end{eqnarray}
which shows that the deviation from the valence quark model result
($S_G=\frac{1}{3}$) is due to the decrease of the probability of the
'bare' valence quark state ($Z_Q<1$) \cite{SMS}.

We now discuss our regularization scheme. Since the above expressions
for the quark distributions involve loop integrals with one of the
LC momentum components fixed, it is clear that we need a
regularization scheme which can be formulated in terms of LC
momenta. Two such schemes which have been discussed extensively in
ref. \cite{LIGHT} are the Lepage-Brodsky (LB) or invariant mass scheme
\cite{LB} , and the
transverse cut-off (TR) scheme \cite{TRAN}. The basic graphs which are 
regularized
in both schemes are the $qq$ and ${\overline q} q$ bubble graphs 
$\Pi_s=\Pi_{\pi}$, the quark-diquark bubble graph $\Pi_N$ and the
quark self energy $\Pi_Q$, either for the case that all internal
momentum components are integrated out or one of the LC momentum
components is fixed. Concerning the LB scheme, it has been shown 
in detail for the case of
$\Pi_s$ in ref. \cite{LIGHT} that, 
if all momentum components are integrated out, this scheme 
is equivalent to the covariant
3-momentum (or dispersion) cut-off scheme. 
\footnote{By 'covariant 3-momentum cut-off scheme' we mean the procedure
where the 3-momentum cut-off is introduced in the particular Lorentz
frame where the total momentum of the two-body ($qq$, ${\overline q}q$,
or quark-diquark) state is zero (${\bold p}=0$), and the result is 
'boosted' to a general frame. For the graphs $\Pi_{\alpha}\,\,(\alpha=
s,\pi,N,Q)$ considered here, this 'boosting' simply means the
replacement $p_0^2\rightarrow p^2$, and for $\Pi_N$ and $\Pi_Q$
also $p_0 \gamma^0 \rightarrow \fslash{p}$. It is known that this
procedure is equivalent to the dispersion cut-off scheme \cite{HK}.}   
Generally, if the intermediate
state involves particles with masses $m_1$ and $m_2$, the LB cut-off
applied in the frame where the transverse components of the total momentum
are zero (${\bf p}_{\perp}=0$) restricts the invariant 
mass of the state according to
\be
\frac{{\bf k}_{\perp}^2+m_1^2}{x}+\frac{{\bf k}_{\perp}^2+m_2^2}{1-x}
<\Lambda_{LB}^2,
\label{lbcut}
\ee
where $x$ and $1-x$ are the fractions of the total momentum component $p_-$
carried by the two particles.
The LB regulator $\Lambda_{LB}$ is related to the 3-momentum cut-off 
$\Lambda_3$ by
$\Lambda_{LB}=\left(\sqrt{m_1^2+\Lambda_3^2}+\sqrt{m_2^2+\Lambda_3^2}\right)$.
For the case of $\Pi_s=\Pi_{\pi}$ we have $m_1=m_2=M_Q$, and the value of
$\Lambda_3$ is determined as usual by requiring that $f_{\pi}=93 MeV$.
In the case of the graphs $\Pi_N$ ($\Pi_Q$) we have $m_1=M_Q$, while
$m_2$ is the mass parameter in the dispersion representation of
$\tau_s$ ($\tau_{\pi}$). In order not to increase the number of parameters,
we will take the same value of $\Lambda_3$ for all graphs $\Pi_s=\Pi_{\pi}$,
$\Pi_N$ and $\Pi_Q$. 

Concerning the TR cut-off scheme, it
has been discussed in ref. \cite{LIGHT} that the use of this scheme
requires a mass renormalization procedure, since the basic self
energies $\Pi_{\alpha}\,\,\,(\alpha=s,\pi,N,Q)$ involve also
logarithmically divergent longitudinal momentum $(k_-)$ integrals, which are
not affected by the TR regularization prescription 
$|{\bold k}_{\perp}|<\Lambda_{TR}$. In this scheme one has therefore
to impose the pion, the diquark and the nucleon masses as renormalization
conditions rather than to relate them to the
parameters $g_{\pi}$, $g_s$ and $r_s$ via the pole conditions. 
For example, if we impose
the condition $1+2g_s\Pi_s(M_D^2)=0$ for some fixed $M_D$, the t-matrix
$\tau_s$ in
eq. \refb{tau} can be re-written in the renormalized form 
$\tau_s=2i/(\Pi_s(k^2)-\Pi_s(M_D^2))$, which is formally 
independent of $g_s$ and free of 
divergencies due to the longitudinal momentum integration. In the
calculation using the TR cut-off we will impose the same value of $M_D$
as obtained in the calculation using the LB cut-off.  
        
\section{Numerical results}

In both the LB and TR regularization schemes
we use $M_Q=400 MeV$ for the constituent quark mass, and determine
the cut-off so as to reproduce $f_{\pi}=93 MeV$. This gives 
$\Lambda_3=593 MeV$ for the equivalent 3-momentum cut-off in the LB
scheme, and $\Lambda_{TR}=407 MeV$ in the TR scheme. In the LB scheme,
we then obtain $g_{\pi}=6.92 GeV^{-2}$ and $r_s=g_s/g_{\pi}=0.727$ from the 
requirements
$m_{\pi}=140 MeV$ and $M_N=940 MeV$, respectively, and the resulting
scalar diquark mass becomes $M_D=600 MeV$. \footnote{The current quark mass
obtained from the gap equation is $m=5.96 MeV$ in the LB scheme.}
As we explained earlier, in the TR
scheme we use the same value $M_D=600 MeV$, and rewrite the t-matrices
$\tau_{\pi}$ and $\tau_s$, which are needed to calculate the distribution
functions, in terms of $m_{\pi}$ and $M_D$ such that they
become independent of $g_{\pi}$ and $r_s$. 

Our results for the valence and sea quark distributions are shown in
figs. 4-7 both for the LB and the TR cut-off scheme. As we have explained 
in sect.1, in order to make contact to the
empirical distributions extracted from the measured structure
functions, we have to evolve our calculated distributions from
the low energy scale $\mu^2=Q_0^2$ to the value $\mu^2=Q^2$ where 
empirical parametrizations are available. For this $Q^2$ evolution
we use the computer
code of ref. \cite{COMP} 
to solve the DGLAP equation in the next-to-leading order. (For the
$Q^2$ evolution we use $N_f=3,\,\,\Lambda_{QCD}=250 MeV.$) 
We will compare our evolved
distributions to
the parametrizations of ref.\cite{MRS} for $Q^2=4 GeV^2$. Both the
calculated and the empirical distribution functions refer to the
$\overline{MS}$ renormalization and factorization scheme. 
The value of $Q_0^2$
is treated as a free parameter which is determined so as to reproduce
the overall features of the empirical valence quark distributions
at $Q^2=4 GeV^2$. In this way we obtain a value
of $Q_0^2=0.16 GeV^2$, i.e; $Q_0$ is equal to our constituent quark mass
$M_Q$.   
    
Let us first discuss the valence quark distributions shown in figs. 4
and 5. Although we do not show the results of the pure valence quark model
(no pions),
we note that the input distributions at $\mu^2=Q_0^2$ shown here 
are softer than in the case without pionic cloud effects, that is, the pionic
effects reduce the peak heights of the valence quark distributions and 
increase their support at low $x$. The integral over the input distributions
shows that at $\mu^2=Q_0^2$ the valence quarks carry $92\,\% \,(87\,\%)$
of the nucleon's LC momentum for the case of the LB (TR) cut-off. The rest
is carried by the sea quarks. This
reduction of the peak heights due to pionic effects has a beneficial 
effect on the overall
behaviour of the valence quark distributions, although it is 
insufficient in particular for the $d$ quark in the LB scheme. The input
distributions are still rather stiff even when pionic effects are taken into
account, which necessitates the use of
a low value of $Q_0^2$ in order to approach the empirical distributions
via the $Q^2$ evolution. We can expect some improvement concerning this
point in a full Faddeev calculation, since in the present static 
approximation the size of the nucleon is too small, corresponding to momentum
distributions which are too stiff. 

In the LB cut-off scheme, the input distributions
are zero for large (and also very small) values of $x$, and therefore the
output distributions (at $Q^2=4 GeV^2$) show a too strong variation
with $x$ compared to the empirical ones. On the contrary, for the TR
cut-off the input distributions are non-zero in the whole region of $x$,
which leads to a smoother behaviour of the output distributions. 
\footnote{For the TR cut-off, the input distributions show a sharp
increase when $x$ becomes very close to $1$. Since the computer code
used for the $Q^2$ evolution \cite{COMP} requires an input distribution
which vanishes for $x=1$, we artificially modified it for $x$ very close
to $1$ such that it goes like $(1-x)^n$ with some power $n$. ($n=10$
was used in the actual calculation.)}
This is
the same feature as noted in ref. \cite{LIGHT} for the quark distribution
in the pion, and indicates that for phenomenological applications the TR
cut-off seems to be superior over the LB cut-off. On the other hand,
as we have explained earlier, the shortcoming of the TR cut-off scheme
is that the diquark mass must be treated as a free parameter since in
this scheme mass renormalizations are necessary in order to get finite
results.  

In our calculation, the difference between the valence $u$ and $d$ quark
distributions reflects the scalar diquark correlations in the proton: 
Since the $d$ quark appears inside the diquark and not as a spectator
quark (see eq.\refb{gen}), its distribution is given by the convolution 
of two distributions (eq.\refb{conv}), which is more concentrated at low 
values of $x$ compared to the spectator quark distribution. This is
in agreement with the behaviour shown by the empirical distributions,
and this observation was in fact one of the motivations to introduce
diquark degrees of freedom also into the bag model description of the 
nucleon structure functions \cite{FLAV}. 

We now turn to the antiquark distributions shown in figs. 6 and 7.
As in the case of the valence quark distributions, the TR cut-off
scheme leads to an overall better agreement with the empirical
distributions than the LB scheme. The enhancement of ${\overline d}$
over ${\overline u}$ is clearly seen both in the input and the output
distributions. Since the numerical value of the probability of the
quark state without pion cloud is $Z_Q=0.84$ ($Z_Q=0.83$) for the
LB (TR) cut-off scheme, the Gottfried sum \refb{gs} becomes 
$S_G=0.262$ ($S_G=0.257$), compared to the experimental value 
\footnote{It has been shown that the Gottfried sum is almost unchanged
by the $Q^2$ evolution \cite{KUM,ROSS}.} reported
by the NM collaboration \cite{EXP} $S_G=0.235 \pm 0.026$. Our results
for the difference $f_{{\overline d}/P}-f_{{\overline u}/P}$ are
shown in fig. 8. We see that with our value of $Q_0^2$ which has been
chosen such as to reproduce the overall behaviour of the valence quark
distributions, the calculated difference is smaller than the empirical
one for intermediate values of $x$, but larger for small $x$. Concerning
the ratio ${\overline u}/{\overline d}$, there are also data from Drell-
Yan processes \cite{DY} which give ${\overline u}/{\overline d}=0.51 \pm 0.04 
\pm 0.05$ at $x=0.18$, compared to our calculated value of $0.68\,\, (0.70)$ 
at $Q^2=4 GeV^2$ for the LB (TR) cut-off. This, too, shows that the
observed flavor asymmetry of the Dirac sea is larger than our calculated
one in this range of $x$.  
  
\section{Summary and outlook}

In this paper we used the framework of the relativistic Faddeev equation
in the
NJL model to calculate the quark LC momentum distributions in the nucleon.
As a first step towards a full Faddeev calculation, we used the nucleon
vertex functions obtained in the simple static approximation to the
Faddeev kernel, and included pionic cloud effect approximately 
using the familiar
convolution formalism. We can summarize our results as follows: First, we
have shown a method based on a straight forward Feynman diagram evaluation, 
which we believe to be best suited for the calculation of the distribution
functions in the Faddeev framework. Besides being simple and straight
forward, the method also guarantees the validity of the number and 
momentum sum rules from the
outset. Second, we discussed two regularization schemes which can be
formulated in terms of LC coordinates, and which preserve the
number and momentum sum rules. Based on our numerical results we 
concluded that for the description of the momentum distributions in the
nucleon the transverse momentum cut-off scheme is superior
over the invariant mass regularization scheme, which is similar to
the situation found previously for the momentum distribution in the pion. 
Third, we
have shown that the resulting distribution functions reproduce the
overall behaviours of the empirical ones if the low energy scale for
the $Q^2$ evolution is taken to be about the same as the constituent
quark mass ($400 MeV$ in our calculation). Such a low value is
required since the input valence quark distributions calculated in our 
model are rather stiff, although the pionic cloud effects served to
soften them compared to the pure valence quark results. In this respect,
a full Faddeev calculation which gives larger and more realistic nucleon
radii, as well as the inclusion of higher mass diquark channels (axial
vector diquark channel) are expected to improve the situation. We have
also shown that the value for the Gottfried sum obtained in
this simple calculation is in basic agreement with the experimental one.

The formulation and results of this work can be used as a basis for
at least the following three extensions: First, one could easily use the
present framework to calculate the quark spin distributions, provided that
the axial vector diquark $(J=1)$ channel is also taken into account. 
Second, one should use the full
Faddeev vertex functions to calculate the quark momentum distributions. As we
have noted in sect.1, the most convenient way for this purpose might
be first to go to the moment space and then to reconstruct the distribution
functions.  
The third extension concerns the case of finite density: 
For a finite density calculation the full Faddeev framework seems to be
intractable, and approximations like the static approximation
used in this paper might be unavoidable. For this purpose, however,
it is necessary first to construct an equation of state for nuclear matter
based on the Faddeev (quark-diquark) picture of the single nucleon, 
similar to the Guichon equation of state \cite{GUI} which is based on 
the MIT bag picture of the single nucleon. The construction of such an 
equation of state and its applications are now under consideration
\cite{FUT}.

\vspace{1cm}
{\sc Acknowledgements}

The authors would like to thank M. Miyama and S. Kumano for the
computer program used for the $Q^2$ evolution (ref. \cite{COMP}).
One of the authors (W.B.) is grateful to A. W. Thomas, A. W. Schreiber, 
K. Suzuki and K. Tanaka for discussions on the nucleon structure function.


\newpage

\section*{Figure captions}

\begin{enumerate}

\item {Graphical representation of the quark LC momentum distribution
in the Faddeev framework. The single (double) line denotes the quark
propagator (scalar diquark t-matrix), the hatched circle stands for the quark-
diquark vertex function, and the operator insertion denoted by
a cross stands for $\gamma^+ \delta(k_- - p_- x)(1\pm \tau_z)/2$ for the
$U(D)$ quark distribution. The second diagram
stands symbolically for those 2 diagrams obtained by inserting the
cross into both particle lines in the diquark. The quark-diquark vertex
contains the isospin operator $\tau_2$. 
The diagram where
the operator insertion is made on the exchanged quark is not shown
here since it does not contribute in the static approximation.}

\item{Feynman diagrams which have to be evaluated in addition to those
shown in fig.1 due to the presence of the pion cloud. Here the
dashed line indicates the ${\overline q}q$ t-matrix in the pionic
channel, and the other lines are as in fig.1. The second diagram
stands symbolically for those 2 diagrams obtained by inserting the
cross into both the $q$ and ${\overline q}$ lines in the pion. 
The dots indicate all
remaining diagrams where the pion line is attached to a quark in the
diquark.}

\item{Graphical representation of the quark distribution within an
on-shell parent quark. The meaning of the lines is as in figs.1 and 2.
The quark spinor $\Gamma_Q(p)$ is associated with the incoming and
outgoing quark lines in all diagrams. The operator insertion in the 
first diagram stands for $\gamma^+ \delta(p_- - p_- x)(1\pm \tau_z)/2$, 
and in the 
other diagrams for $\gamma^+ \delta(k_- - p_- x)(1\pm \tau_z)/2$ for the
u(d) quark distributions. 
The third diagram
stands symbolically for those 2 diagrams obtained by inserting the
cross into both the $q$ and ${\overline q}$  lines in the pion.} 

\item {LC momentum distributions of the valence $u$-quark in the proton, using
the LB (solid lines) and TR (dashed lines) regularization scheme.
The lines associated with $Q_0^2=0.16 GeV^2$ show the NJL model results, 
and the lines associated with $Q_0^2=0.16 GeV^2$ show the results
obtained by the QCD evolution in next-to-leading order from 
$Q_0^2=0.16 GeV^2$ to $Q^2=4 GeV^2$, using $\Lambda_{QCD}=0.25 GeV$
and $N_f=3$. The dotted line shows the parametrization for $Q^2=4 GeV^2$ 
obtained in ref. \cite{MRS} by analyzing the experimental data.}

\item{LC momentum distributions of the valence $d$-quark in the proton.
For explanation of the lines, see the caption to fig. 4.}

\item{LC momentum distributions of the ${\overline u}$-quark in the proton.
For explanation of the lines, see the caption to fig. 4.}

\item{LC momentum distributions of the ${\overline d}$-quark in the proton.
For explanation of the lines, see the caption to fig. 4.}

\item{The difference of ${\overline d}$ and ${\overline u}$-quark 
momentum distributions in the proton.
For explanation of the lines, see the caption to fig. 4.}
 
\end{enumerate}

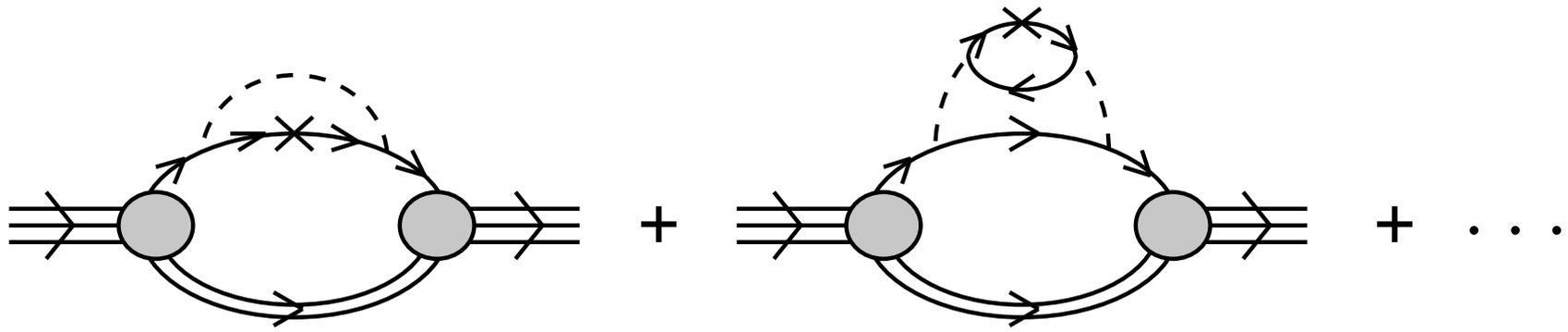
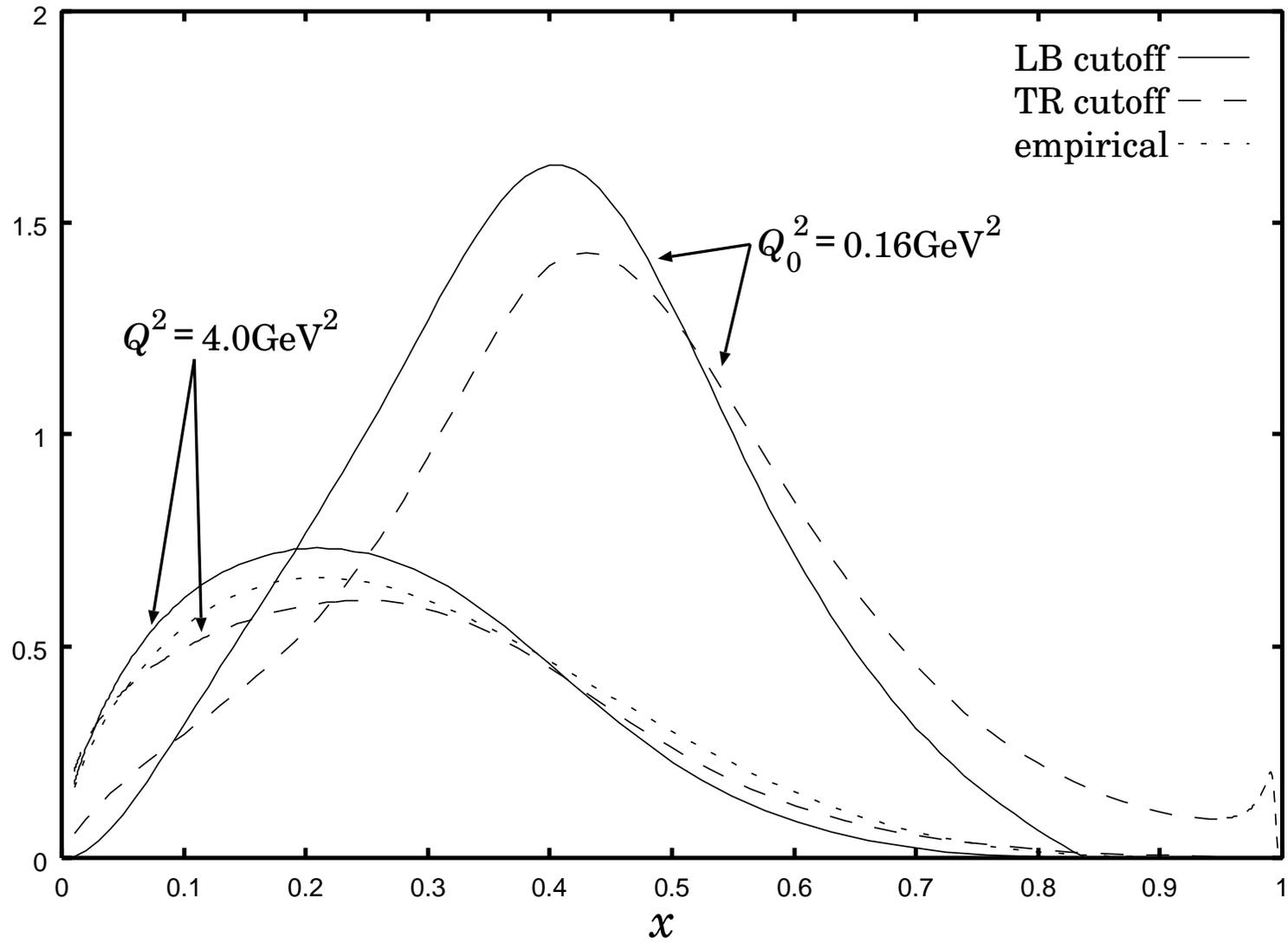
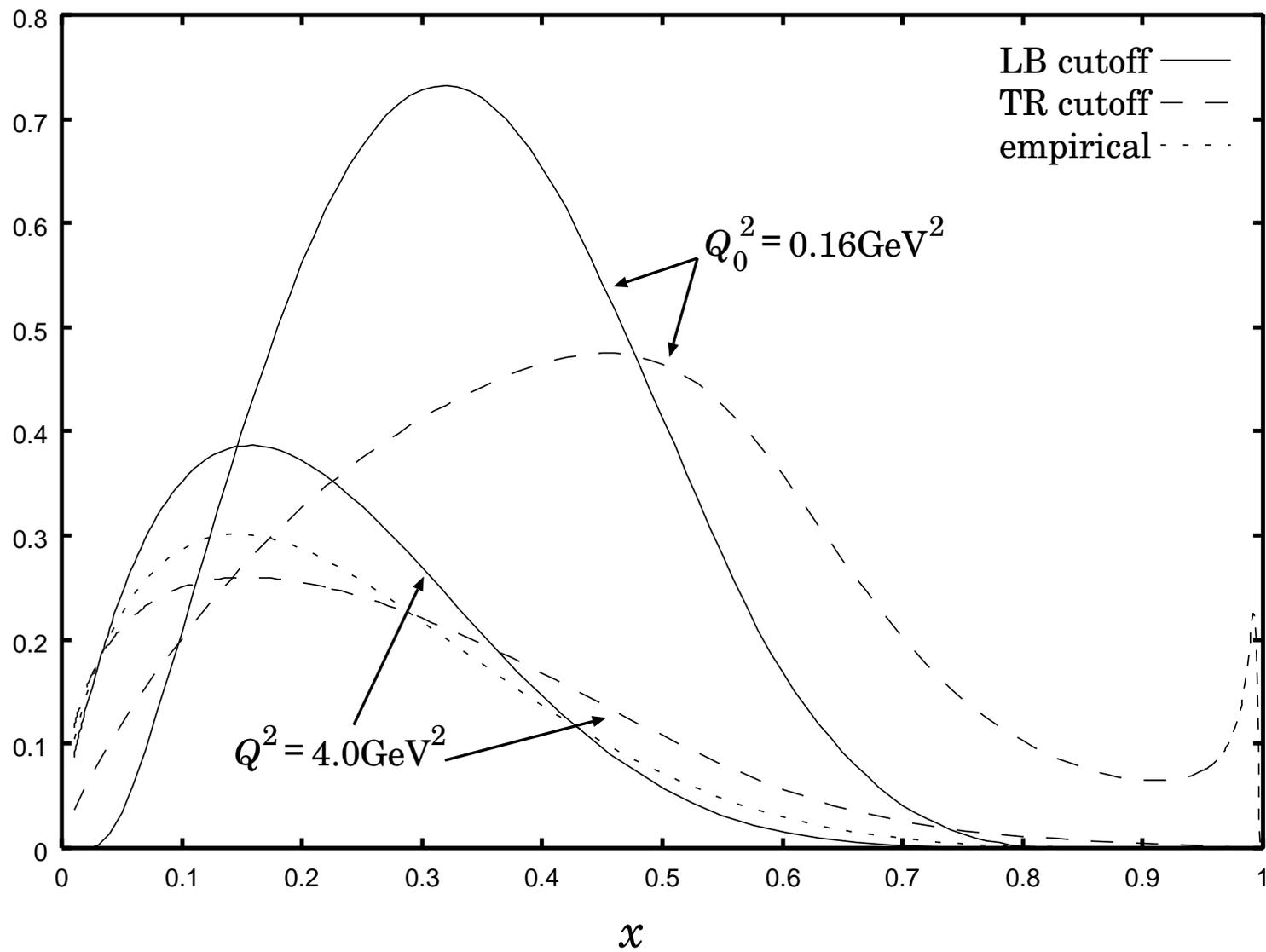
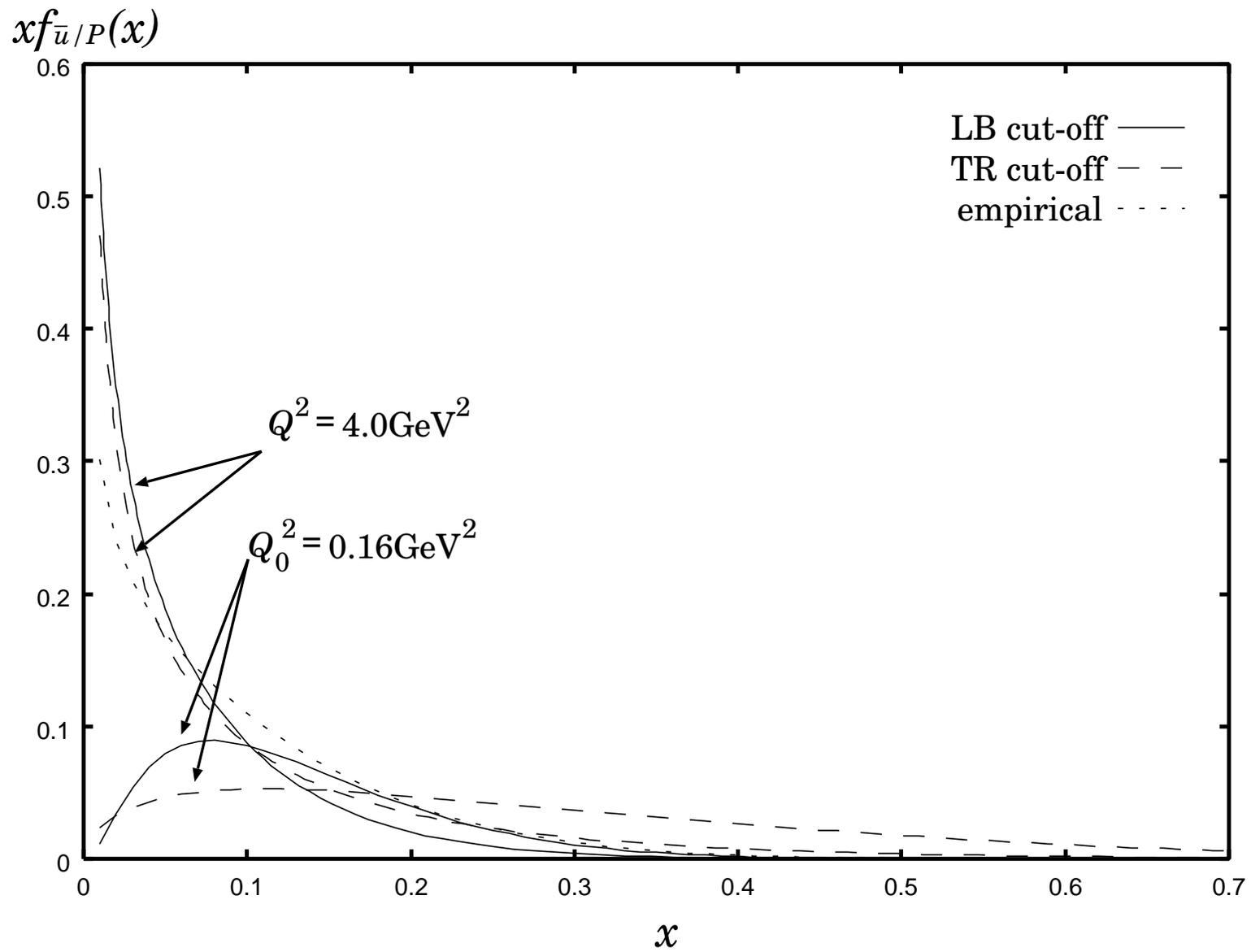
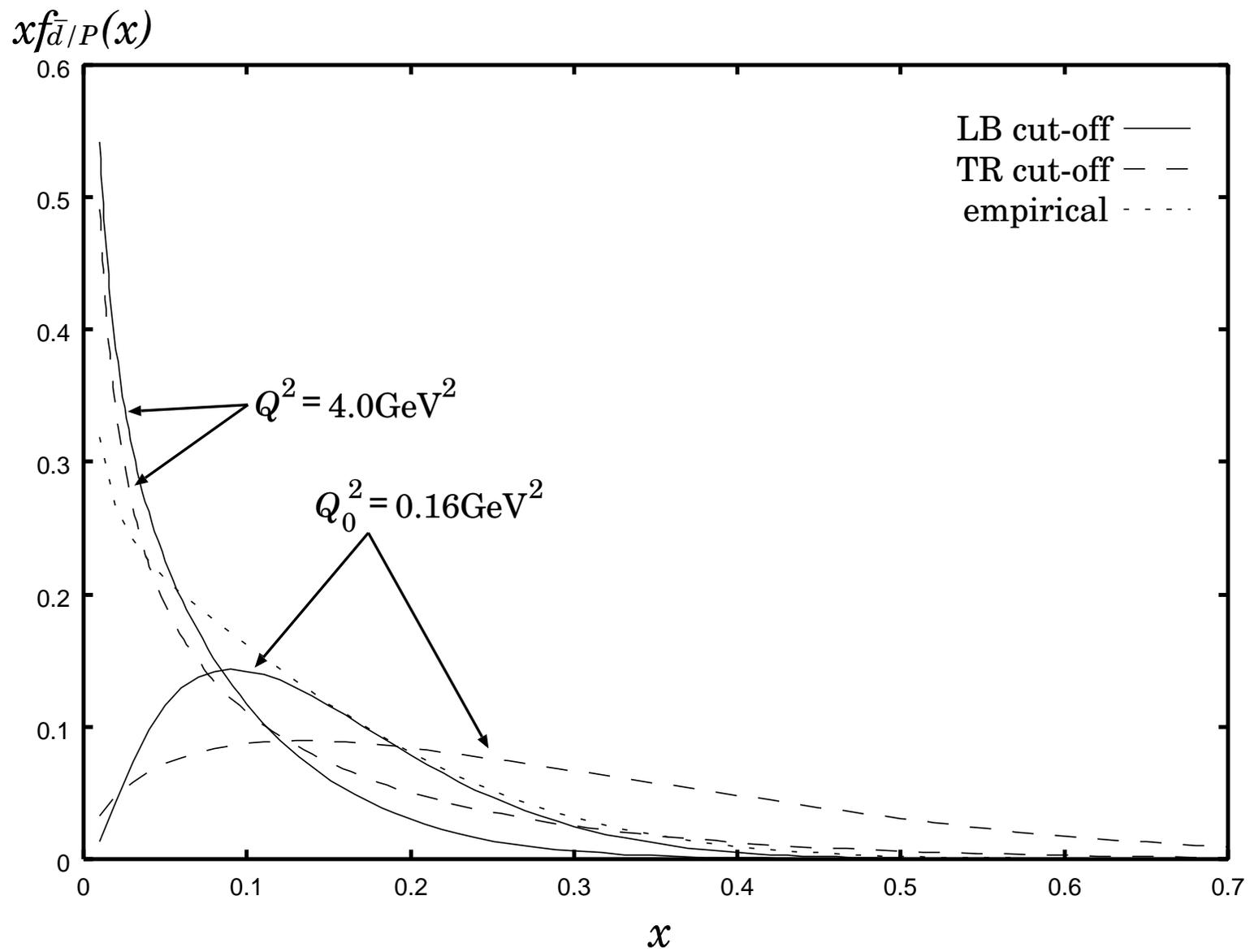
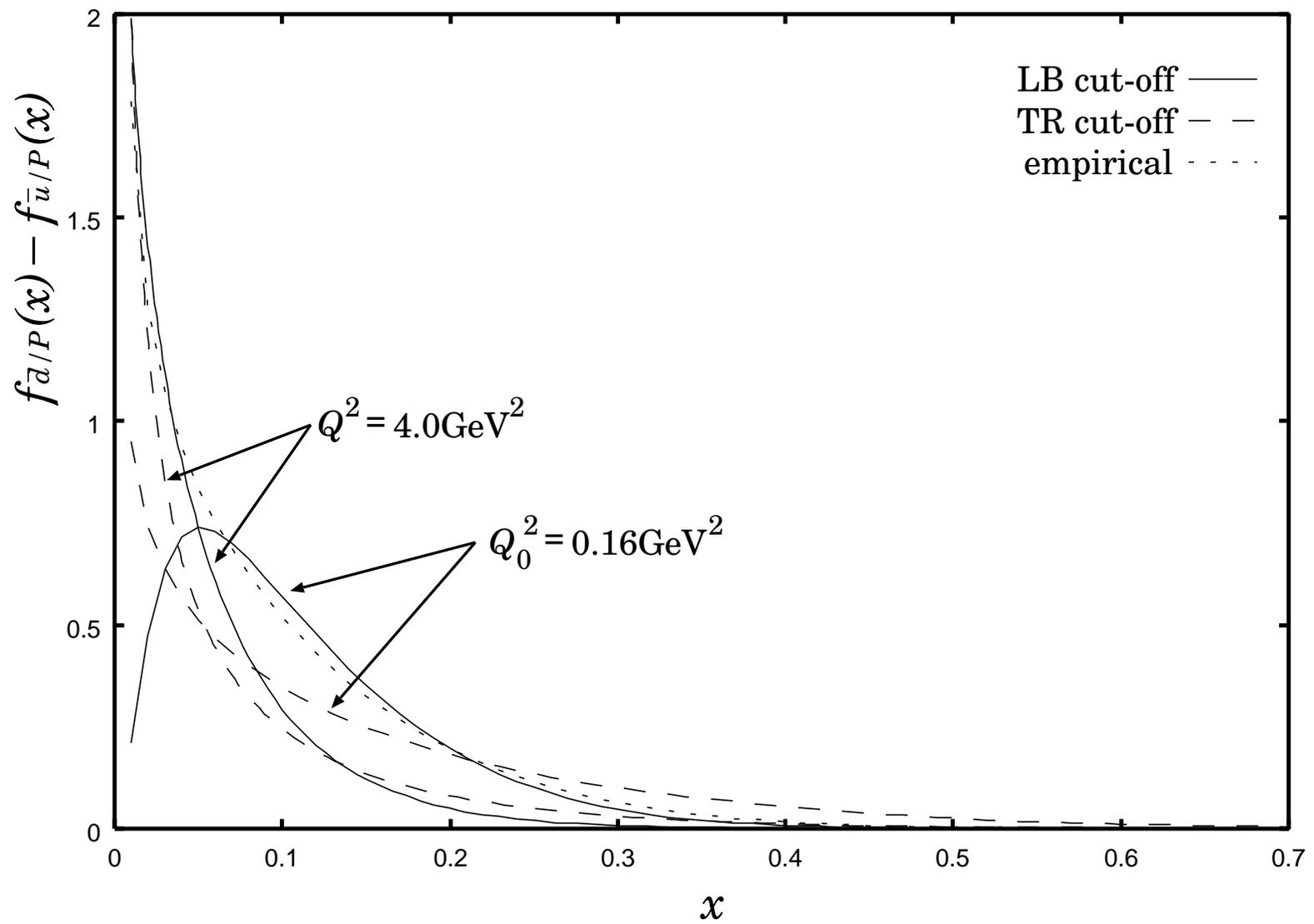
\end{document}